# Electrically Tunable and Reconfigurable Topological Edge State Lasers

Ruizhe Yao[1†], Hang Li[2†], Bowen Zheng[2], Sensong An[2], Jun Ding[3], Chi-Sen Lee[1], Hualiang Zhang[2]* and Wei Guo[1]*

[1]*Department of Physics and Applied Physics, University of Massachusetts Lowell, Lowell, Massachusetts, USA*
[2]*Department of Electrical & Computer Engineering, University of Massachusetts Lowell, Lowell, Massachusetts, USA*
[3]*School of Information and Science Technology, East China Normal University, Shanghai, China*

† These authors contributed equally to this work.
*hualiang_zhang@uml.edu, wei_guo@uml.edu

**Abstract-**We report an actively tunable topological edge mode laser in a one-dimensional Su-Schrieffer-Heeger (SSH) laser chain, where the SSH chain is realized in an electrically-injected Fabry-Perot (FP) laser chain. A non-Hermitian SSH model is developed to investigate the SSH laser chain with tunable active topological defect. We theoretically demonstrate topological edge mode phase transition in the SSH laser chain. The phase transition manifested in the tight binding SSH laser chain is observed experimentally and agreed well with the theoretical predictions. Finally, by electronically tuning the gain and loss, a lossy topological mode is obtained, and lasing is experimentally observed in the lossy time-reversal configuration. This work presents a versatile platform to investigate novel concepts, such as topological mode, for main stream photonic applications.

## 1. Introduction

Topological insulators (TIs) introduce a new form of matter, which can support the flow of electrons on their surface but are insulator in their interior. Unlike surface states in normal insulators, the surface states in TIs are robust against perturbation and defects in the bulks [1-6]. Such tantalizing characteristics address the impurity issues in conventional bulk topology materials, and spark interests in the field of topological photonics [7-9]. Thus, it has drawn large interests in the investigation of topological phase transitions and edge-states in optical devices, such as lasers, that are unaffected by local perturbations and fabrication defects [10-16]. Most of such work has been focused on optical structures and devices with passive components. By incorporating active components in the photonic topological structures, it becomes possible to achieve lasing in these topological edge-states. The resulting topological edge state lasers can exhibit low lasing threshold compared to the conventional bulk state lasers and are robust against any fabrication defects and local defects. Topological edge state lasers have been recently demonstrated in several systems[17-22], however, all of them are shown in optically pumped platforms (bulky and slow) which can largely limit the further exploration and applications of these unique topological lasers. In addition, it has been paid large attention in studying the interplay between non-Hermiticity and photonics topology devices. The optical non-Hermitian system is brought into attention by the notion of parity–time (PT) symmetry, where several new novel phenomena and optical devices have been proposed and demonstrated based on the PT symmetry in non-Hermitian systems [23-30]. In this context, topological lasers become an ideal platform owing to the inherent non-Hermiticity in laser structures. In this report, we demonstrate lasing in topological edge state in one-dimensional Su-Schrieffer-Heeger (SSH) Fabry-Perot (FP) laser chain, which contains gain and loss waveguides and an active gain topological defect. The active topological defect is employed to enhance the density of state (DOS) of the topological edge state and ensure lasing at the lowest threshold compared to other bulk modes. Phase transition between the topological edge state and bulk state lasing is shown experimentally. Finally, by electrically tuning the gain and loss in the SSH laser chain, the new configuration is a time-reversal of the original one and the topological defect mode becomes a lossy mode, and, by increasing the gain, lasing is also obtained in the SSH chain in the time-reversal configurations.

## 2. Topological Laser Design

To achieve a photonics topological edge mode laser, we design a photonics topological insulator structure consisting of 7 pairs of coupled Fabry-Perot (FP) waveguide SSH dimers. Figure 1(a) shows the schematic of the dimer chain, where $C_1$ and $C_2$ represent the coupling efficiency between the gain (orange) and loss (blue) waveguides, and $C_1 > C_2$. The trivial and non-trivial topological states are achieved by strong and weak coupling in the dimers consisting of gain and loss, respectively. A topological defect in the chain is introduced by removing the loss waveguide in the middle dimer (4$^{th}$ pair). As a result, the designed active SSH chain consists of 13 coupled ridge waveguide lasers with the width of 3 µm. The laser active region is composed of 7-layer InAs quantum-dots (QDs) sandwiched between Al$_{0.4}$Ga$_{0.6}$As cladding layers. Figure 1(b) shows the heterostructures of the InAs QD lasers grown by molecular beam epitaxy (MBE) system. In the laser chain, the top p+-GaAs contact layer and p-Al$_{0.4}$Ga$_{0.6}$As cladding layer are etched for electrical isolation between each waveguide. The coupling strengths, $C_1$ and $C_2$, in the dimers are controlled by the width of the isolation trenches, 1 µm and 2 µm, respectively. All the gain and loss waveguides are interconnected allowing simultaneous and fully electronic control of all the gain and loss waveguides.

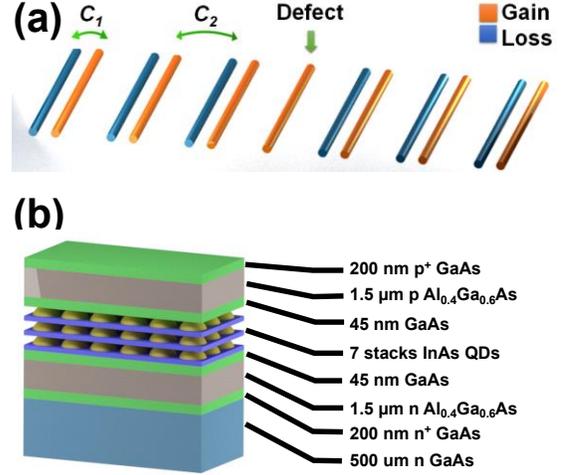

FIG. 1: (a) Schematic of the dimer chain, where $C_1$ and $C_2$ represent the coupling efficiency between the gain (orange) and loss (blue) waveguides, and $C_1 > C_2$. (b) heterostructures of the InAs QD lasers grown by molecular beam epitaxy.

## 3. Non-Hermitian SSH Model

A one-dimensional SSH model is first employed to investigate the passive Hermitian dimer chain without gain and loss [31]. The SSH model studies two sites per unit cell with different coupling efficiency $C_1$ and $C_2$. It is well understood that, in SSH mode, the edge states depend on the configuration of the unit cell. The edge state exists only when the smaller coupling, $C_2$, is located at the edges [31, 32]. In contrast, there is no edge state if the chain is terminated by a dimer with larger coupling efficiency, $C_1$. These two different scenarios can be identified by the winding number in Brillouin zone (BZ) SSH model [33]:

$$W_h = \frac{1}{2\pi} \int_{BZ} dk \frac{\partial \psi(k)}{\partial k} \quad (1)$$

where k is the Bloch wavenumber within the first BZ, and the winding number is corresponding to the Zak phase ($\psi_{ZAK}$) divided by $\pi$ [34]. It has been shown that, when $W_h=1$, the system exhibits trivial topological phase without edge states, in contrast, if $W_h=0$, the system transits into non-trivial topological phase with two edge states presented [35]. As a result, in the explored topological configuration, the defect waveguide, $4^{th}$ pair dimer with gain waveguide only, is located at an interface between two structures that have different topological invariants, $W_h = 0, 1$. Thus, there must be a topological interface state residing at zero energy at the defect waveguide. Szameit *et. al.* has already investigated such topological waveguide configuration containing passive components and shown the topological edge mode appearing in the center defect while preserving the PT symmetry [36]. In this work, we employ the dimer chain containing gain and loss and an active, gain, defect to achieve topological laser. It has been argued that, if an active defect is introduced, the topological edge state can be enhanced by supplementing the topological protection with non-Hermitian symmetries [37]. In this context, a non-Hermitian SSH model is employed to explore the density of state (DOS) of the topologically zero energy mode (see supplemental information). As shown in Figure 2, the DOS of the zero-energy mode is largely enhanced by including an active defect compared to the passive defect case. It is worth noting that the DOS enhancement can be obtained not only with the active defect but also with an absorptive lossy defect due to the symmetry of the Hamiltonian [37].

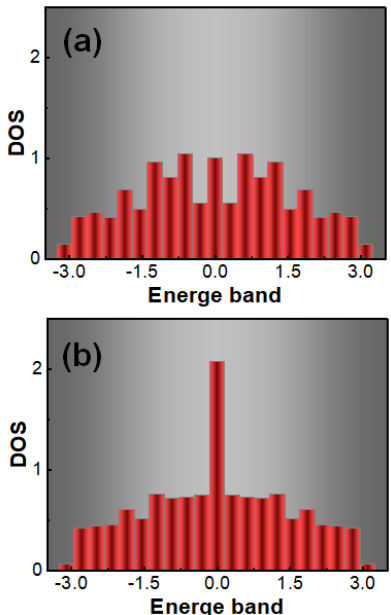

FIG. 2: Density of state (DOS) of the topologically zero energy mode without (a) and with (b) active defect, the DOS of the zero-energy mode is largely enhanced by including an active defect compared to the passive defect case.

In addition, it is also of large interests to investigate the PT symmetry in the proposed gain/loss non-Hermitian SSH system. As discussed by Szameit *et. al.* [36], the PT symmetry is always preserved in the passive dimer chain with a neutral defect [36]. In our case, the active gain defect explicitly breaks the PT symmetry, preventing the Hamiltonian from commuting with the PT operators, [PT, H] ≠ 0. As shown in Figure 2, the topologically induced property can further enhance the mode competition. The topological defect state is predominately located in the gain defect waveguides and is isolated from loss waveguides affecting all other modes in the system, which can reduce the lasing thresholds of the topological mode, even in presence of structural disorder. Figure 3 shows the eigenvalue diagrams and corresponding optical field distributions at different gain, $g$, values, where a dimensionless variable $\gamma = g/C_1$ is defined in this work. It is found that the edge mode exhibits complex eigenvalues, with gain, as long as the system gain is non-zero and all the optical field are predominately located in the defect gain waveguide in the SSH system and symmetrically distributed around the defect waveguide. By increasing the gain, other bulk modes show complex eigenvalue, shown in the Figure 3(b) and 3(c). The dynamics of the proposed SSH laser is theoretically investigated. As shown in Figure 3(d), in phase I, only the edge mode exhibits complex eigenvalue with gain, and the SSH laser chain is lasing at single topological mode. The corresponding dispersion relation of the lasing edge mode and optical field distribution are plotted in Figure 3(a). By increasing the gain, in phase II, one trivial bulk mode is transited into a phase with complex eigenvalue and starts to lase alongside with edge mode. Figure 3(b) shows the dispersion relation of the lasing edge mode with complex eigenvalue and the corresponding optical field distribution of the lasing bulk mode. Finally, with further increase of the gain, other bulk modes show complex eigenvalues and the gains of these bulk modes are comparable with the edge mode. Therefore, the edge mode becomes less dominant and this in turn results in a more uniform mode intensity profile across the SSH chain under investigation. The dispersion relation and optical field distribution of the second bulk mode under this case are shown in Figure 3(c).

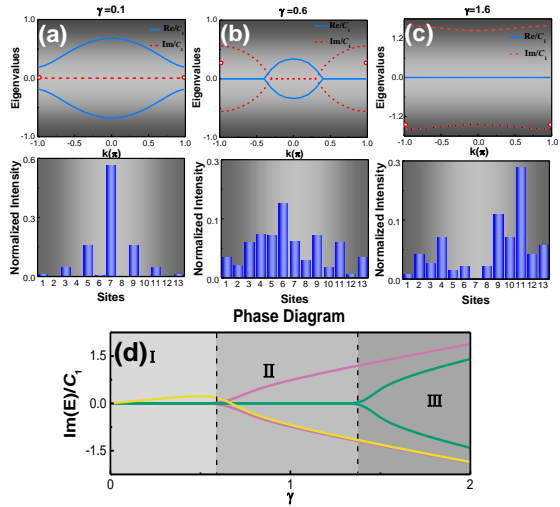

FIG 3: Complex eigenvalue diagram of the SSH laser chain in Phase I (a), II (b) and III (c); the insets below show the corresponding optical intensity distributions from theoretical predication; (d) Phase diagram of the complex SSH chain.

Finally, Figure 4(a) plots the band diagram of topological system with neither gain nor loss. It is clearly seen that the system has a topological band gap with a protected edge state at $k = \pi$, while other zero energy state in the band gap at $k \neq \pi$ are trivial modes due to the odd number of sites in the system [38]. In addition, Figure 4(b) shows the evolution of complex eigenvalues with the γ varied from 0.1 (blue dots) to 0.3 (red dots). It is found that when the system gain level is increased the gain increment of the topological mode is ~10 times the ones of trivial modes. This different mode evolution behavior explicitly shows the mode selection between the non-Hermitian topological mode and trivial bulk modes.

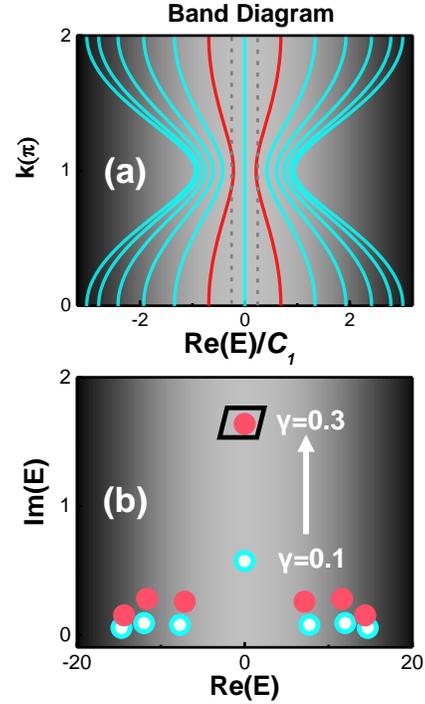

FIG 4: (a) Band diagram of topological system with neither gain nor loss. It is clearly seen that the system has its topological properties without the presence of gain. (b) Evolution of complex eigenvalues with the γ varied from 0.1 (blue dots) to 0.3 (red dots). It is found that when the system gain level is increased the gain increment of the topological mode is ~10 times the ones of trivial modes.

## 4. Results

In the experimental demonstration, the gain/loss SSH chain is achieved by electrically biasing the gain and loss waveguides at different levels. The bias current of the loss waveguides is maintained at 0 mA, which corresponds to a loss of ~ 35 cm$^{-1}$ in the loss waveguide, and the gain waveguide bias current is varied from 0 to 1000 mA to electrically tune the gain in the gain waveguides. Figure 5 shows the light-current (L-I) characteristic of the SSH coupled waveguide laser chain. It is observed that the lasing threshold current is ~ 600 mA. The inset shows the electroluminescence (EL) spectrum of the laser chain. The lasing wavelength at 1.32 μm is obtained.

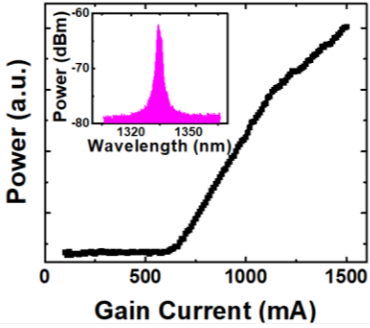

FIG 5: Light-current (L-I) characteristic of the SSH laser chain. It is observed that the lasing threshold current is ~ 600 mA. Inset: Electroluminescence (EL) spectrum of the laser chain. The lasing wavelength at 1.32 µm is obtained

The near-field pattern of the SSH laser chain is measured to investigate the dynamics of the topological mode lasing and mode selections (see supplemental information). Shown in Figure 6, the near-field patterns from the laser chain are measured at various gain bias currents. When the gain bias current is small, 630 mA, the SSH laser chain only lases at the topological mode in the center defect and the mode profile exhibits intensity maximum at the center defect. When the bias current is increased to 700 mA, the first trivial bulk mode obtains enough gain and transits to lasing mode, while the topological mode is still dominating in the near-field profile. The SSH laser system evolves to phase II. With further increment of the bias current to 770 mA, more trivial bulk modes start to obtain enough gain and eventually transform to lasing modes, which leads to a more uniform near-field profile across the laser chain, and the system is transited into phase III. The experimental observation shows clear phase transitions of the topological mode laser and agrees well with the theoretical predications, Figure 3(d).

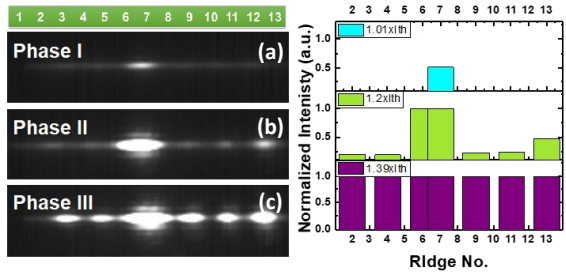

FIG 6: Near-field patterns measured at various gain bias currents. When the gain bias current is small, 630 mA (a), the SSH laser chain only lases at the topological mode in the center defect and the mode profile exhibits intensity maximum at the center defect, site 7. When the bias current is increased to 700 mA (b), the first trivial bulk mode obtains enough gain and transits to lasing mode, while the topological mode is still dominating in the near-field profile. The SSH laser system evolves to phase II. With further increment of the bias current to 770 mA (c), more trivial bulk modes start to obtain enough gain and eventually transform to lasing modes, which leads to a more uniform near-field profile across the laser chain, and the system is transited into phase III.

Finally, we also investigate a topological configuration with a lossy topological defect in the SSH chain. In the experiment, owing to the versatility of the invented electrically injected system, the lossy topological defect is simply achieved by interchanging the gain and loss waveguide bias conditions. The resulting Hamiltonian in the new configuration is equivalent to applying the time-reversal operator to the Hamiltonian of the original system. This new configuration still features the same topological configuration as in the case discussed before, but now with a lossy defect at the topological interface (see supplemental information). The corresponding dispersion relations and phase diagram show that the topological mode exhibits net loss at small gain and lasing is prohibited in Phase I (see supplemental information). With the increment of system gain level, both the topological and bulk modes start to obtain gain, and lasing of topological and bulk modes occurs in Phase II. Figure 7(a) shows the lasing optical field distribution of the topological mode and bulk mode at $|\gamma|=1.2$. It is worth noting that at higher gain, unlike in the previous configuration, the topological mode profile exhibits minimums at the center defect waveguide sites. This can be clearly seen from the topological mode profile evolution at different gain levels (see supplemental information). The lossy topological configuration is also demonstrated experimentally, and the corresponding near-field patterns are measured. Shown in Figure 7(b), instead of lasing at single topological mode, both topological and bulk modes lase at the same time and the measured near-field pattern represents the superposition of these modes. The near-field pattern of the original gain defect topological laser at higher gain is shown as well to illustrate the relative position of each site. The lasing threshold current of the lossy topological defect configuration is ~700 mA obtained from the L-I measurement. The higher lasing threshold compared to the gain defect topological laser is due to that the lasing can only occur at Phase II instead of Phase I in this case. Finally, it is worth nothing that, in addition to the demonstration of the reconfigurability of the topological SSH chain, the lossy topological defect configuration is of interests for applications such as optical modulators and detectors, where the enhanced absorption, in Phase I, is beneficial.

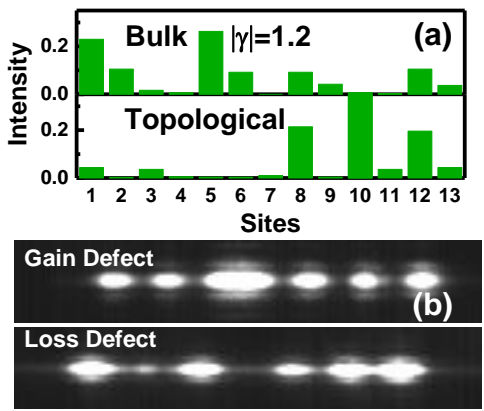

FIG 7: (a) Optical field distribution of topological mode and bulk mode; (b) near-field characteristics of the gain and lossy defect topological lasers.

### 5. Conclusion
In summary, we have demonstrated an electrically-injected topological edge mode laser in a SSH chain with an active gain defect. The explored laser system is described by non-Hermitian SSH model. Phase transitions between lasing of topological and bulk modes are observed. The reconfiguration of the system into a lossy defect topological laser is also investigated theoretically and experimentally. The experimental results are in good agreement with theoretical predictions. This work demonstrates a versatile electrically tunable platform and provides new perspectives in understanding some of the fundamentals in non-Hermitian topological systems.

### 6. Acknowledgements

R.Y. and W.G. acknowledge funding support from Massachusetts Clean Energy program. H.L., B.Z., S.A., and H.Z. gratefully thank funding support provided by NSF under award CMMI-1661749. The authors also acknowledge fabrication facility support by the Harvard University Center for Nanoscale Systems funded by the National Science Foundation under award 0335765.

### 7. Author contributions
W.G. and H.Z. conceived the ideas. R.Y. and C.S.L. contributed to the growth and fabrication the devices. R.Y. and H.L. performed device characterizations. H.L. modeled the device. H.L., R.Y., B.Z., S.A., J.D., H.Z., and W.G. analyzed the experimental data. W.G. and H.Z. supervised and coordinated the project. H.L., R.Y., H.Z., and W.G. wrote the manuscript. All authors contributed to technical discussions regarding this work.

### 8. Competing financial interests
The authors declare no competing financial interests.